\documentclass[10pt]{article}

\usepackage{epsf,amsfonts,amssymb,epsfig,amsmath,amsthm}
\usepackage{multirow}
\usepackage{dsfont}
\usepackage{bbold}

\usepackage{latexsym}
\usepackage{euscript}
\usepackage{mathrsfs} 
\usepackage{enumerate}
\usepackage{dsfont}

\usepackage{amssymb,amsmath,mathbbol,mathrsfs}
\usepackage[usenames,dvipsnames]{color}
\usepackage[linktocpage=true]{hyperref}
\usepackage{xcolor}
\usepackage{stmaryrd}
\usepackage{authblk}
\usepackage{framed}
\usepackage{empheq}
\usepackage{slashed}
\usepackage{hyphenat}
\usepackage{cite}

\usepackage{mdframed}

\renewcommand{\text}[1]{#1}

\newcommand{\be}{\begin{equation}}
\newcommand{\ee}{\end{equation}}
\newcommand{\ben}{\begin{displaymath}}
\newcommand{\een}{\end{displaymath}}
\newcommand{\bea}{\begin{eqnarray}}
\newcommand{\eea}{\end{eqnarray}}
\newcommand{\bean}{\begin{eqnarray*}}
\newcommand{\eean}{\end{eqnarray*}}
\newcommand{\ba}{\begin{array}}
\newcommand{\ea}{\end{array}}
\newcommand{\bi}{\begin{itemize}}
\newcommand{\ei}{\end{itemize}}

\def\1f{f_1^{1/2}}
\def\2f{f_2^{1/2}}
\def\4f{f_4^{1/2}}

\def\linebreak{\hfil\break}

\begin{document}
\begin{titlepage}

\vfill



\begin{center}
  \baselineskip=16pt
  {\Large\bf  O Nature, Where Art Thou?}\footnote{A version of this article was published as ``Unifying gravity and quantum theory requires better understanding of time'',  Nature \textbf{645}, 32-34 (2025)}
 \vskip 1.5cm

Fay Dowker\\
    \vskip .6cm
           \begin{small}
     \textit{ 
      {Abdus Salam Centre for Theoretical Physics,\\ Blackett Laboratory, Imperial College, Prince Consort Road, London, SW7 2AZ, UK}\\\vspace{5pt}
              {Perimeter Institute, 31 Caroline Street North, Waterloo ON, N2L 2Y5, Canada}
               }
             \end{small}                       
\end{center}

\vskip 2cm
\begin{center}
\textbf{Abstract}
\end{center}
\begin{quote}
Where does what happens happen in a quantum system?  The standard textbook formulation of quantum mechanics provides a strange, imprecise and yet successful-in-practice answer to this question. In the struggle to unify our understanding of gravity with quantum theory, though, the textbook answer no longer suffices, and an alternative approach is needed. The Feynman Sum Over Histories approach provides an alternative that is particularly suited to quantum gravity because the Sum Over Histories and General Relativity are built on the same fundamental concepts of `event' and `history'. 
 \end{quote}

\vfill
\end{titlepage}


\newpage

\section*{The Copenhagen D\'etente}

Quantum mechanics is our most successful physical theory.  Created to account for atomic phenomena, quantum mechanics has by now a vast range of application far beyond the atomic realm, from predictions of the abundances of the light elements created a few minutes after the Big Bang to understanding the properties of semiconductor materials that are the basis of advanced information processing technology. Quantum mechanics is successful not only in its vast scope but also in its exquisitely accurate predictions such as the measured value of the magnetic moment of the electron, predicted by Quantum Electrodynamics to an accuracy of one part in ten thousand billion. Our best current understanding of the fundamental constituents of matter, the Standard Model of particle physics, is a quantum theory and the statement that every physical system is, fundamentally, a quantum system has no known counter-evidence.\\

 In the case of {gravity}, however, nearly a century of effort has not resulted in a stable consensus even about the most promising grounds on which to build a theory of quantum gravity. Why is quantum gravity proving more challenging than other quantum theories? The reasons lie partly in the lack of definitive observational phenomenology to guide us and partly in the character of gravity that makes it different from all other physical theories. In this article I diagnose ``where the shoe pinches’’—Albert Einstein's phrase—in quantum gravity and describe one way forward based on Richard P. Feynman’s alternative vision for quantum mechanics.\\

To begin at the beginning with the famous breakthrough made by Werner Heisenberg on the island of Helgoland in 1925 \cite{Heisenberg:1925zz}, it is clear in retrospect that the truly novel and startling concept introduced by Heisenberg is that of {transitions} of the atom from one quantum state to another quantum state, transitions  \textit{that do not occur in physical three-dimensional space}. That breakthrough set up a conflict between the new rules of quantum prediction on the one hand and the familiar, historically successful conceptual framework of physical goings-on occurring in  three-dimensional space on the other.  The question, \textit{If atomic transitions do not occur in 3d space,  where do they occur?} was not explicitly raised by Heisenberg in 1925 but it was explicitly answered soon afterwards. The location, the whereabouts of these transitions was identified and formalised mathematically by Paul Dirac and others as being the so-called `Hilbert space' of the quantum system. The Hilbert space of the system comprises all conceivable quantum states of the system and takes the form of what is known in mathematics as a vector space: think of a quantum state as an arrow pointing in a particular direction in a space with many many dimensions.\\

The transitions in Hilbert space from one quantum state vector to another quantum state vector  affect and are affected by what occurs in physical 3d space by means of special interactions between the quantum system and instruments that exist in 3d space. Such interactions are known as {measurements} and the standard quantum rules make predictions about the outcomes of measurements registered by the measuring instruments in 3d space. To make these predictions, the goings-on in Hilbert space must be coordinated somehow with the goings-on in 3d space and this coordination is achieved using a concept of synchronized time. \\
 
Time passes in Hilbert space and the quantum state of the quantum system changes, evolves, moves in Hilbert space as time passes. The motion of the quantum state in Hilbert space is governed by a law of quantum physics, the Schr\"odinger equation, and is a type of rotation: the angle through which the quantum state vector  rotates in Hilbert space is proportional to the time that passes. Time also passes for the measuring instruments and the physicists in physical 3d space.  The situation is like the round in  the  long-running BBC comedy radio show ``I'm Sorry I Haven't a Clue" in which each celebrity contestant has to sing along with a recording of a song, then continue to sing while the recording is muted. When the recording is unmuted again the contestant wins points if they are still singing in synch with the recording. Seriously, it's funnier than it sounds. The recording playing is like the quantum state evolving in Hilbert space and the singer is like the measuring instrument in three-dimensional space. The period during which the singer cannot hear the recording is like the period during which the measuring instrument is not making a measurement. The moment of unmuting is like the moment of measurement. In the game the singer can get out of synch with the recording but  in quantum mechanics the synchronisation is always perfect. The time that has passed in 3d space for the measuring instrument is always in perfect synch with the time that has passed in Hilbert space marked by the angle the state vector has rotated through.  This perfect synchronisation is so fundamental to quantum mechanics that, when working in the theory, the same symbol $t$ is used  for time in 3d physical space and for time in Hilbert space.\\

By means of the concepts of measurement and of times  synchronized between Hilbert space  and  physical 3d space, Heisenberg and his colleagues in Copenhagen established a scientifically successful  d\'etente between the two notions of the whereabouts of physical goings-on: the evolution of the quantum state of the quantum system occurs in Hilbert space, while scientific predictions are about measuring instrument events that occur in physical 3d space. The last one hundred years have proven that one can be an extremely successful quantum physicist by accepting and working with this strange duality of location. When the quantum system is \textit{gravity}, however,  the Copenhagen d\'etente cannot hold. \\

Gravity is not like other physical systems. Our best theory of gravity is General  Relativity or GR for short. In GR the physical entity that is the subject of the theory is spacetime itself. Spacetime is a physical four-dimensional fabric with geometrical structure that bends, warps, ripples, carries energy and has its own laws of motion as precise and as experimentally successful as Newton’s laws of mechanics.  In GR, spacetime graduates from being a fixed stage on which what happens happens,  to being both the stage and a dynamic actor in its own right  in reality's play. Gravity's physical status is different from other systems’. If particles, for example, are absent from a region of spacetime, there are still physical goings-on in that region, including warping and rippling spacetime goings-on. But if {spacetime} is absent, there cannot be any particles, cannot be any electromagnetic radiation, cannot be any anything because there's no-where and no-when for them to be. Nothing is external to spacetime, every physical thing that exists in GR either {is} spacetime or is {in} spacetime. \\

The Copenhagen requirement for a physical measuring instrument in physical space external to the quantum system, is incompatible with the quantum system being spacetime. Also, the character of physical time in GR stymies the synchronization that the Copenhagen d\'etente requires. In GR, physical time passes individually for each particle or body or measuring instrument along its own unique, individual worldline in spacetime.  These individual physical worldline times cannot even be synchronised with each other and so certainly cannot be synchronised with the time that passes in Hilbert space as demanded by the Copenhagen rules. For quantum gravity  the d\'etente collapses and an alternative approach is needed.

\section*{The Feynmanian approach, an alternative foundation}

In 1985 the particle physicist Richard P. Feynman gave a series of public lectures about Quantum Electrodynamics (QED)  \cite{FeynmanQED}, the theory for which he shared the 1965 Nobel Prize for physics. In the lectures, Feynman describes an approach to quantum theory in which there is no strange duality of location because quantum states evolving in Hilbert space are absent from the approach. Feynman bases his explanations of the phenomena of QED on alternative concepts:  events and histories in spacetime \cite{ Feynman:1948ur}.
These concepts of event and history are also fundamental in GR  and starting in the 1980’s two physicists, James B. Hartle 
\cite{Hartle:1989,Hartle:1991bb} 
and Rafael D. Sorkin 
\cite{Sorkin:1987cd, Sorkin:1994dt, Sorkin:1995nj} 
have sought to build on this conceptual unity between Feynman’s approach and GR  and  to develop Feynman’s approach into an alternative, stand-alone foundation for quantum theory suitable for quantum gravity.\\

How does the Feynmanian approach based on events and histories work? An {event} is something that can happen. It’s useful to have concrete examples in mind, so think of rain on a particular date in Bengaluru India, or the death of a cat in a particular box at a particular time in Dublin Ireland.  The key properties of events are that an event has a spacetime location and that an event either happens or doesn’t happen. Beforehand, it is uncertain whether a given event will or will not happen, but afterwards there is no uncertainty: it either rained or it did not, the cat either died or did not.  A {history} for an event is a detailed account of one way in which the event could happen. There are many histories for a given event because the detail of the histories can be different for the same event. For the rain event for example the exact number of raindrops and the positions where they fall will differ from history to history. In Feynmanian quantum physics, a history contains much more detail than just numbers of raindrops. A history contains as much detail as the physical theory at hand can provide, such as exact trajectories through spacetime of all the particles that comprise the whole system in quantum mechanics. Predictions in Feynmanian quantum physics are predictions about events and the probability of a particular event happening is calculated like this: one finds all the detailed histories for that event and for each of the histories the theory provides a complex number. Add up the complex numbers for all the histories for the event, square the modulus of the sum and the result is the Feynman Probability of the event.  \\

This Feynmanian approach based on events and histories is often called the Sum Over
Histories, for obvious reasons. The Sum Over Histories approach reproduces, for all practical purposes, the Copenhagen probabilities of events that are outcomes of measurements by measuring instruments, and it does so without reference to quantum states evolving in Hilbert space. \\

Now, everyone knows that  quantum mechanics is strange. So where is the strangeness in the Sum Over Histories approach? There is no strange duality of location: everything that happens, microscopic or macroscopic, happens in spacetime. There are no strange, special measurement interactions between measuring instruments in 3d space and quantum states in Hilbert space: there are just histories in spacetime of measuring instruments and systems being measured, all together, treated in the same way, on the same footing. \\

 But the quantum strangeness must be there somewhere, like a ruck in the carpet that is smoothed out in one spot only to appear in another. In the Sum Over Histories approach, the ruck in the quantum carpet is that the Feynman Probabilities  do not behave as probabilities should for all events. If one calculates the Feynman Probabilities for macro-events like rain or cat deaths or measurement outcomes registered on measuring instruments, then the probabilities work just fine. But the Feynman Probabilities for micro-events, such as where the individual elementary particles actually go, don’t add up like probabilities should, essentially because a Feynman Probability is not a sum but the square of a sum.  Hartle and Sorkin have differing proposals for the direction in which to push this ruck in the Sum Over Histories carpet: Hartle's  proposal is to restrict  the collection of events that should be considered as physical \cite{Hartle:1992as, Hartle:2006nx} whereas Sorkin's proposal is to hang the interpretation on the concept of ``preclusion'' \cite{Geroch:1984, Sorkin:2006wq, Sorkin:2007uc, Sorkin:2010kg}. It remains an open issue today but we do not need to wait for a resolution to see how the Sum Over Histories approach to quantum theory naturally accommodates gravity. The lesson of General Relativity is incorporated into the Sum Over Histories by recognising that the physical geometric structure of the spacetime substrate of an event is an essential physical component of the event and should therefore also be summed over. Each of the histories summed over in the Sum Over Histories for the Feynman Probability of an event in quantum gravity, then, is a spacetime with matter in it. \\
 
 The Sum Over Histories approach to quantum gravity is a work in progress but its perspective already influenced a successful prediction about cosmology. Using the heuristic of a histories approach to quantum gravity, and the additional hypothesis that events are atomic—i.e. every event is comprised of a finite number of indivisible smallest events—Sorkin predicted, in the late 1980's and early 1990's, that the so-called Cosmological Constant, a controversial quantity in cosmology theory, should be non-zero. And moreover, Sorkin predicted, the Cosmological Constant---often referred to by the Greek capital letter $\Lambda$ or by the placeholder moniker Dark Energy---should actually not be constant at all but should fluctuate throughout the history of the cosmos, with an expected magnitude today of roughly $10^{-120}$ in natural units
\cite{Sorkin:1990bj,Sorkin:1997gi, Ahmed:2002mj}. 
Sorkin’s prediction  was spectacularly verified in 1998 by observations of  distant supernovae which indicated an accelerating cosmic expansion consistent with a present value of $\Lambda$ of the predicted order of magnitude \cite{Riess_1998,Perlmutter_1999}. There are even indications from more recent cosmological data sets that $\Lambda$ may indeed not be constant  \cite{desicollaboration2025desidr2resultsi, desicollaboration2025desidr2resultsii}.\\

From the quantum physics of the atom to the quantum physics of the Cosmological Constant $\Lambda$, we've come a long way and the next one hundred years is likely to be an era in which microscopic and cosmic physics intertwine even further. Our understanding of the fundamental nature of our quantum world may be informed by observations about the largest system of them all, the entire universe.
\section*{Acknowlegements}
The author is supported in part by 
STFC Consolidated Grant ST/X000575/1. Research at Perimeter Institute is supported by the Government of Canada through
Industry Canada and by the Province of Ontario through the Ministry of Economic Development and Innovation.

\bibliography{../Bibliography/refs}
\bibliographystyle{../Bibliography/JHEP}

\end{document}